\documentclass{article}
\usepackage{amsmath}
\usepackage{epsfig}
\usepackage{graphics}
\usepackage{graphicx}
\usepackage{color}

\def\rmd{\mathrm{d}}
\def\rmi{\mathrm{i}}
\def\kv{{\bf k}}
\def\Gv{{\bf G}}
\def\Gvp{{\bf G}'}
\def\rvs{{\bf r}_<}
\def\rvl{{\bf r}_>}
\def\rv{{\bf r}}
\def\rvp{{\bf r}'}
\def\rvpp{{\bf r}''}
\def\rvppp{{\bf r}'''}
\def\rvh{{\bf \hat r}}
\def\rvhp{{\bf \hat r}'}
\def\rvhpp{{\bf \hat r}''}
\def\deps{\rmd \epsilon}
\def\drv{\rmd {\bf r}}
\def\drvh{\rmd {\bf \hat r}}
\def\drvhp{\rmd {\bf \hat r}'}
\def\drvp{\rmd {\bf r}'}
\def\drvpp{\rmd {\bf r}''}
\def\drvppp{\rmd {\bf r}'''}
\def\Rn{{\bf R}^n}
\def\Rnp{{\bf R}^{n'}}
\def\nr{n({\bf r})}
\def\nrp{n({\bf r}')}
\def\rrp{|{\bf r}-{\bf r}'|}
\def\EF{E_{\mathrm{F}}}
\def\lmax{l_{\mathrm{max}}}
\def\lpot{l_{\mathrm{pot}}}
\def\Vproj{\hat{V}}
\def\nproj{\hat{n}}
\def\Nsph{N_{\mathrm{sph}}}
\def\Nat{N_{\mathrm{at}}}
\def\Ts{T_{\mathrm{s}}}

\def\Ts{T_{\mathrm{s}}}
\def\Etot{E_{\mathrm{tot}}}
\def\Ehxc{E_{\mathrm{hxc}}}
\def\Exc{E_{\mathrm{xc}}}
\def\Enn{E_{\mathrm{nn}}}
\def\Gcut{G_{\mathrm{cut}}}
\def\Vext{V_{\mathrm{ext}}}

\begin{document}

\title{Precise Kohn-Sham total-energy calculations
at reduced cost}
\author{Rudolf Zeller \\
Institute for Advanced Simulation \\
Forschungszentrum J\"ulich GmbH and JARA \\
D-52425 J\"ulich, Germany}

\date{}

\maketitle

\begin{abstract}
The standard way to calculate the Kohn-Sham orbitals utilizes an
approximation of the potential. The approximation consists
in a projection of the potential into a finite subspace
of basis functions. The orbitals, calculated with the projected potential,
are used to evaluate the kinetic part
of the total energy, but the true potential is
used to evaluate the interaction energy with the electron
density. Consequently, the Kohn-Sham total-energy expression loses
its stationary behaviour as a functional of the potential. It will be discussed
that this stationarity is important for the calculation of precise total energies
at low computational cost and an approach will be presented that practically restores 
stationarity by perturbation theory.
The advantage of this approach will be illustrated with total-energy results
for the example of a disordered CrFeCoNi high entropy alloy.
\end{abstract}

\section{Introduction} 
\label{sec:intro}
The standard way to solve the Kohn-Sham density-functional equations
is based on the variational principle of the total energy. The Kohn-Sham
orbitals are expanded into a finite set of basis functions and the total
energy is minimized with respect to the expansion coefficients. 
In practice, this means that the Kohn-Sham effective potential, defined
as the sum of external, Hartree and exchange-correlation potentials, 
is approximated by a projection into
a finite subspace of basis functions. This approximation
leads to first order
errors for the orbitals, the density and the potential.
For precise total-energy calculations, these first order errors
do not represent a problem because the Kohn-Sham total-energy expression
can be written in a form \cite{ref:HH95, ref:H98}, which is stationary
with respect to orbitals, density and potential. This means
that these first order errors lead to only
second order errors for the total energy.

Unfortunately, however, the approximation of the potential by
projection into a finite subspace of basis functions
introduces another important error that is associated 
with the evaluation
of the total-energy functional.
While the kinetic part of the total energy is evaluated with the 
projected potential, the true potential is used for
the evaluation of the interaction energy between the potential 
and the electronic density.
As a consequence of the two different potentials,
the total-energy expression loses
its stationarity as a functional of the potential and
total-energy calculations are considerably more expensive than they should be.
It is important to note that this problem is not removed if the quality of
the basis set is improved. The use of more basis functions
or the choice of better basis functions can reduce the size
of the evaluation error, but cannot change its fundamental
unfavourable property that it destroys the stationarity with respect
to the potential.

In present article it is shown how the error arising
from the use of two different potentials can be determined and that
this information can be used to obtain 
corrections which considerably improve the calculated total energies.
Because the corrections must be used only once after
the density-functional self-consistency steps, 
only a small overhead of computing time is required to obtain
precise total energies. For the example of
projection into finite subspaces of spherical harmonics it is shown
that almost as precise total energies are obtained by
perturbation theory.
Because perturbation theory is computationally inexpensive, it can be 
used during the self-consistency
steps to modify the projected potential such that 
density, potential and total energy
are obtained in a consistent manner.

The outline of the article is as follows. In section \ref{sec:te} the total-energy
functional and its stationarity properties are discussed. 
In section \ref{sec:PP} non-local potentials are introduced, which describe projection
into subspaces of spherical harmonics, and it is explained that
for these angular projection potentials
the density and the kinetic energy can be determined numerically practically exactly.
Section \ref{sec:PP} also provides two expressions that
can be used to determine the total-energy
error either from the single-particle
energies or from the interaction energy between density and potential.
In section \ref{sec:num} it is shown how perturbation theory can be used to obtain
precise total energies with considerably reduced computational resources.
Section \ref{sec:summa} contains a summary and an outlook.
In the appendix the evaluation error is discussed for plane wave methods
in order to motivate further work beyond the present article for projection
potentials which correspond to other basis functions than spherical harmonics.
 
\section{Total-energy functional}
\label{sec:te}
The total-energy of Kohn-Sham density-functional theory
is usually written as
\begin{equation}
\label{eq:Etot}
\Etot[\nr] = \Ts[\nr] + \int \drv \Vext(\rv) \nr + \Ehxc[\nr] + \Enn .
\end{equation}
Here $\nr$ is the electron density,
$\Vext(\rv)$ the external potential provided by the nuclei, 
$\Enn$ the interaction energy of the nuclei  
and
\begin{equation}
\label{eq:Ehxc}
\Ehxc[\nr] = \frac{e^2}{2} \int \int \frac{\nr \nrp}{\rrp}
\drv \drvp + \Exc [\nr]
\end{equation}
the sum of Hartree and exchange-correlation energy. 
$\Ts[\nr]$ is the kinetic energy of the non-interacting Kohn-Sham
reference system, which in Rydberg atomic units $\hbar^2 / 2m = 1$ is given by 
\begin{equation}
\label{eq:Ts1}
\Ts [\nr] = \sum_i n_i \int \drv \varphi^{\star}_i (\rv)
\left[ - \nabla^2_{\rv} \right] \varphi_i (\rv) .
\end{equation}
Here $\varphi_i (\rv)$ are the Kohn-Sham orbitals and
$n_i$ are occupation numbers which are one for occupied orbitals and
zero for unoccupied orbitals.
It is well known that $\Etot[n(\rv)]$ is stationary with respect to $\nr$
and that the minimum energy $\Etot^0$ is obtained for the ground-state density $n_0(\rv)$.

Haydock and Heine have shown \cite{ref:HH95, ref:H98} that the functional
(\ref{eq:Etot}) can be generalized into a functional
which is stationary
with respect to density, potential and orbitals.
This functional can be written as
\begin{align}
\label{eq:GEtot}
\Etot[\nr] & =
\sum_i n_i \int \drv \varphi^{\star}_i (\rv)
\left[ - \nabla^2_{\rv} + V(\rv) \right]
\varphi_i (\rv) \\
& - \int \drv V(\rv) \nr + \int \drv \Vext(\rv) \nr + \Ehxc [\nr] + \Enn \nonumber \\
& - \sum_i n_i \epsilon_i \left[ \int \drv \varphi^{\star}_i (\rv) \varphi_i (\rv) - 1 \right]
\nonumber
\end{align}
where Lagrange multipliers
$\epsilon_i$ are used to enforce the normalization of the orbitals.
Variation of (\ref{eq:GEtot}) with respect to the orbitals leads to the Kohn-Sham equation
\begin{equation}
\label{eq:KS}
\left[ - \nabla^2_{\rv} + V(\rv) \right] \varphi_i (\rv) = \epsilon_i \varphi_i (\rv)
\end{equation}
for the orbitals,
variation with respect to the density leads to the standard expression
\begin{equation}
\label{eq:Veff}
V(\rv) = \Vext(\rv) + \frac{\delta \Ehxc [\nr]}{\delta \nr}
\end{equation}
for the effective potential and variation with respect to the
potential leads to the standard expression
\begin{equation}
\label{eq:dens}
\nr = \sum_i n_i \varphi^{\star}_i (\rv) \varphi_i (\rv)
\end{equation}
for the density. If (\ref{eq:KS}-\ref{eq:dens}) are solved
self-consistently, the functional (\ref{eq:GEtot}) gives the
usual density-functional total energy for the ground state. Furthermore, if $\varphi_i(\rv)$,
$V(\rv)$, $\nr$ are approximated, the total-energy error is of second order
in $\delta \varphi_i(\rv)$, $\delta V(\rv)$, $\delta \nr$.

For the example of projection into a finite subspace of spherical harmonics
it has been shown recently \cite{ref:Z15a, ref:Z15b}
that the Green function for the differential equation (\ref{eq:KS})
can be calculated practically exactly.
This means that no approximation must be made for the orbitals and the 
density. Thus, the relevant error arises from $\delta V(\rv)$ while errors arising from 
$\delta \varphi_i(\rv)$ and $\delta \nr$ are negligible.
If, as suggested in \cite{ref:HH95, ref:H98},
the total-energy error is of second order in $\delta V(\rv)$, the calculated 
total energy should rapidly improve with an increasing subspace
of spherical harmonics
and only small subspaces should be necessary for precise results.
This rapid improvement, however, is not
found. This implies that the error is not of second order and that the
stationarity of (\ref{eq:GEtot}) with respect to $\delta V(\rv)$
is not fulfilled. The reason for this deficiency is that 
the second term in the total-energy functional (\ref{eq:GEtot})
must contain the true potential
in order that the condition (\ref{eq:Veff}) is satisfied whereas
the first term of (\ref{eq:GEtot}) contains the approximated
projected potential.

\section{Angular projection potentials}
\label{sec:PP}
The approximation of the potential by a finite number of matrix elements
can be understood as a projection of the potential into a finite
subspace of basis functions. The projected potential is
non-local which complicates the mathematical and numerical treatment
because the single-particle Schr\"odinger equation is transformed from
a differential equation
into an integro-differential equation, formally by the replacement
\begin{equation}
V(\rv) \varphi_i(\rv) \Rightarrow
\int V(\rv,\rvp) \varphi_i(\rvp) \drvp .
\end{equation}
This means that the local
potential, which acts as multiplicative factor, is replaced by the non-local potential
which acts as an integral operator.

For projection into subspaces of spherical harmonics the non-local potential
has the form
\begin{equation}
\label{eq:V0}
V(\rv,\rvp) = \frac{\delta(r-r')}{r^2} V(r,\rvh,\rvhp)
\end{equation}
with
\begin{equation}
\label{eq:V1}
V(r,\rvh,\rvhp) =
\sum_{LL'}^{\lmax} Y_{L} (\rvh) V_{LL'} (r) Y_{L'} (\rvhp) ,
\end{equation}
where $L=(l,m)$ is a combined index for the angular quantum numbers.
For systems which are described by a sum of such angular projection potentials,
centred at the atomic positions $\Rn$, the integro-differential
Schr\"odinger equation can be solved \cite{ref:Z13} by a generalization
of the Korringa-Kohn-Rostoker Green-function (KKR-GF) method.
The important advantage of using non-local projection potentials
instead of local potentials is that the matrix elements
\begin{equation}
\label{eq:VLL}
V_{LL'} (r) = \int \drvh \int \drvhp 
Y_L(\rvh) V(r,\rvh,\rvhp) Y_L(\rvhp) 
\end{equation}
for the projection potential
are exactly zero for $l > \lmax$
and for $l' > \lmax$ because of the orthogonality of the spherical harmonics.
In contrast to this, a local potential $V(\rv)$ is described
by a infinite number of matrix elements
\begin{equation}
\label{eq:VLL0}
V_{LL'} (r) = \int \drvh Y_L(\rvh) V(\rv) Y_{L'}(\rvh)
\end{equation}
and thus can be treated only approximately. An important consequence of
the exact description of the non-local angular projection potentials by a
finite number of matrix elements is that 
the density and the Green function can be expressed
by a finite number of terms
with analytically known dependence on the angular variables \cite{ref:Z15a, ref:Z15b}.
Only the radial dependence must be treated numerically. With the present computer
capabilities this can be done practically exactly. This means
that the precision of the total energies is essentially determined by
the single parameter $\lmax$ and that an increase of $\lmax$, which improves the
the agreement between the projection potential and the true effective potential,
should increase the precision.

\subsection{Green function}
\label{sec:GF}

The Green function $G( \rv , \rvp; \epsilon)$ for the
integro-differential Schr\"odinger equation 
can be obtained by solving the integral equation
\begin{equation}
\label{eq:Dyson}
G( \rv , \rvp; \epsilon) =
g ( \rv , \rvp; \epsilon) + 
\int \drvpp g ( \rv , \rvpp; \epsilon)
\int \drvppp V ( \rvpp, \rvppp)
G( \rvppp, \rvp;\epsilon) .
\end{equation}
Here $g ( \rv , \rvp; \epsilon)$ is the analytically known Green function
of free space (with zero potential), which is defined by
the differential equation
\begin{equation}
\label{eq:gfree}
\left[ - \nabla^2_{\rv} - \epsilon \right]
g( \rv , \rvp; \epsilon) = - \delta (\rv - \rvp) .
\end{equation}
As shown in detail in \cite{ref:Z13},
for a sum of angular projection potentials of type (\ref{eq:V1}),
centred at the atomic positions,
the Green function can be written as
\begin{align}
\label{eq:GF}
G(\rv+\Rn, \rvp + \Rnp ; \epsilon ) =
\delta_{nn'}
& \sum_L S^n_L(\rvl; \epsilon) R^n_L(\rvs; \epsilon) \\
+ & \sum_{L L'} R^n_L(\rv;\epsilon)
G_{LL'}^{nn'} (\epsilon) R^{n'}_{L'}(\rvp;\epsilon)
\nonumber
\end{align}
with matrix elements
\begin{equation}
\label{eq:aDyson}
G^{nn'}_{LL'}  (\epsilon) =g^{nn'}_{LL'} (\epsilon) +
\sum_{n''} \sum^{}_{L''L'''} g^{nn''}_{LL''} (\epsilon)
t^{n''}_{L''L'''} (\epsilon) G^{n''n'}_{L'''L'} (\epsilon) 
\end{equation}
that can be obtained by solving this linear matrix equation either
directly or by iterative techniques \cite{ref:TZB12}.
Here $g^{nn''}_{LL''} (\epsilon)$ are the analytically known matrix elements
of the free-space Green function $g ( \rv , \rvp; \epsilon)$.
Note that the symbols $\rv$ and $\rvp$
in (\ref{eq:GF}), as well as below in (\ref{eq:G1}) and
(\ref{eq:VnG}-\ref{eq:DEtot0}),
denote coordinates originating at the atomic positions $\Rn$
(with $\rvs$ and $\rvl$ being defined as the ones with larger and smaller length),
while elsewhere they denote coordinates in all space. 

The functions $R^n_L(\rv)$ and $S^n_L(\rv)$ are simple 
products of spherical Bessel or Hankel functions
with spherical harmonics for $l > \lmax$, while
for $l \le \lmax$ they are given by
finite expansions in spherical harmonics \cite{ref:Z15a}.
As a consequence of this fact,
the angular dependence of the Green function
(\ref{eq:GF}) is also analytically known \cite{ref:Z13} as
\begin{equation}
\label{eq:G1}
G(\rv+\Rn, \rvp + \Rnp ; \epsilon ) = \sum_{LL'}^{} 
Y_L (\rvh) Y_{L'} (\rvhp)
{\cal{G}}^{nn'}_{LL'} (r, r' ; \epsilon ) ,
\end{equation}
a formula which will be used in the next subsection.

\subsection{Kinetic energy}
\label{sec:Ekin}

For the evaluation of the total-energy functional
one needs the density
\begin{equation}
\label{eq:densG}
n(\rv) = - \frac{2}{\pi} \lim_{\rvp \rightarrow \rv} {\mathrm{Im}}
\int_{-\infty}^{\EF} \deps G(\rv,\rvp;\epsilon + \rmi 0^{+})
\end{equation}
and the kinetic energy
\begin{equation}
\label{eq:Ts0}
\Ts [\nr] = - \frac{2}{\pi} \int \drv \lim_{\rvp \rightarrow \rv}
{\mathrm{Im}} \int_{-\infty}^{\EF} \deps
\left[ - \nabla^2_{\rv} \right] G(\rv,\rvp;\epsilon + \rmi 0^{+}) .
\end{equation}
$\EF$ is the Fermi level and $0^{+}$
an infinitesimally small positive quantity which indicates
that the integral must be performed in the complex $\epsilon$ plane
just above the real $\epsilon$ axis.
The factor two in these equations, which is valid for non-spin-polarized systems,
simplifies the notation.
Generalization to spin polarized systems is straightforward.
With the density of states
\begin{equation}
\label{eq:dos}
n(\epsilon) = - \frac{2}{\pi} \int \drv
\lim_{\rvp \rightarrow \rv} {\mathrm{Im}} G(\rv,\rvp;\epsilon + \rmi 0^{+}) ,
\end{equation}
the kinetic energy can be expressed as
\begin{align}
\label{eq:Ts2}
\Ts [\nr]
& = \int_{-\infty}^{\EF} \deps \epsilon n(\epsilon) \\
& - \frac{2}{\pi} \int \drv
\lim_{\rvp \rightarrow \rv}
{\mathrm{Im}} \int_{-\infty}^{\EF} \deps
\left[ - \nabla^2_{\rv} - \epsilon  \right] G(\rv,\rvp;\epsilon + \rmi 0^{+}) .
\nonumber
\end{align}
Here insertion of 
\begin{equation}
\label{eq:diffDyson}
\left[ - \nabla^2_{\rv} - \epsilon \right]
G( \rv , \rvp; \epsilon) = - \delta (\rv - \rvp) 
- \int \drvpp V ( \rv, \rvpp)
G( \rvpp, \rvp;\epsilon) ,
\end{equation}
which is obtained from (\ref{eq:Dyson})
by application of the operator
$\left[ - \nabla^2_{\rv} - \epsilon \right]$
and by use of (\ref{eq:gfree}), leads to
\begin{equation}
\label{eq:Ts2a}
\Ts [\nr]
= \int_{-\infty}^{\EF} \deps \epsilon n(\epsilon)
+ \frac{2}{\pi} \int \drv 
\int \drvpp V (\rv,\rvpp)
{\mathrm{Im}} \int_{-\infty}^{\EF} \deps
G (\rvpp,\rv;\epsilon + \rmi 0^{+})
\end{equation}
because the delta function as a real quantity
gives no contribution to the imaginary part.

The last term in (\ref{eq:Ts2a}) can be simplified
by writing the integrals over the spatial coordinates as 
a sum of integrals over the atomic cells and by inserting
(\ref{eq:V0}), (\ref{eq:V1}) and (\ref{eq:G1}). This leads to
\begin{align}
\label{eq:VnG}
\frac{2}{\pi} \sum_n^{} \int_n \drv & \int_n \drvpp V^n (\rv,\rvpp) 
{\mathrm{Im}} \int_{-\infty}^{\EF} \deps
G (\rvpp+\Rn,\rv+\Rn;\epsilon + \rmi 0^{+}) \\
& = \frac{2}{\pi} \sum_n^{} \int_n \drv
\int_n \drvpp \frac{\delta(r-r'')}{r''^2}
\sum_{LL'}^{\lmax} Y_{L} (\rvh) V^n_{LL'} (r'') Y_{L'} (\rvhpp) \nonumber \\
& \qquad \times
\sum_{L''L'''}^{}
Y_{L''} (\rvhpp) Y_{L'''} (\rvh)
{\mathrm{Im}} \int_{-\infty}^{\EF} \deps
{\cal{G}}^{nn}_{L''L'''} (r'', r ; \epsilon + \rmi 0^{+})
\nonumber \\
& = - \sum_n^{} \int_n r^2 \rmd r \sum_{LL'}^{\lmax} V^n_{LL'} (r)
n^n_{L'L} (r) .
\nonumber
\end{align}
For the last result
the integrals over $r''$, $\rvhpp$ and $\rvh$ were evaluated by the help of
the delta function and by the orthogonality of the spherical harmonics and the
integral over $\epsilon$ was written in terms of the matrix elements
\begin{equation}
\label{eq:nLL}
n^n_{LL'} (r) = - \frac{2}{\pi} {\mathrm{Im}}
\int_{-\infty}^{\EF} \deps {\cal{G}}^{nn}_{LL'} (r, r ; \epsilon + \rmi 0^{+})
\end{equation}
of the density. 
The use of (\ref{eq:VnG}) in (\ref{eq:Ts2}) leads to the final result
\begin{equation}
\label{eq:Ts3}
T_{\mathrm{s}} [\nproj(\rv)] =
\int_{-\infty}^{\EF} \deps \epsilon \nproj(\epsilon)
- \sum_n^{} \int_n r^2 \rmd r \sum_{LL'}^{\lmax} \Vproj^n_{LL'} (r)
\nproj^n_{L'L} (r)
\end{equation}
for the kinetic energy,
where the hat indicates
that the potential, the density of states and the density are the ones
calculated self-consistently using
(\ref{eq:V1}) with a finite value of $\lmax$.

Expression (\ref{eq:Ts3}) is
an important result. First, it shows how the kinetic energy for non-local potentials
of type (\ref{eq:V1}) can be evaluated without numerical differentiation using the
Laplace operator. Second it shows that only $L$, $L'$ components 
with $l \leq \lmax$ and $l' \leq \lmax$ are needed in the second term.
Because also only these
$L$, $L'$ components are needed to evaluate the first term
as explained in section 4.3 of \cite{ref:Z13},
the computing time increases as $\Nat^3 (\lmax+1)^6$,
where $\Nat$ is the number of atoms in the system,
which provides considerable savings for low values of $\lmax$.
Third, in comparison to
\begin{equation}
\label{eq:Ts4}
T_{\mathrm{s}} [\nr] =
\int_{-\infty}^{\EF} \deps \epsilon n(\epsilon)
- \sum_n^{} \int_n r^2 \rmd r \sum_{LL'}^{} V^n_{LL'} (r)
n^n_{L'L} (r) ,
\end{equation}
which is the kinetic energy for the true effective potential, expression (\ref{eq:Ts3})
shows that the use of a finite number of matrix elements changes
the kinetic energy in two ways, implicitly by the dependence of
the density of states, the potential and the density on $\lmax$ and explicitly by
limiting the sums to terms with $l \leq \lmax$ and $l' \leq \lmax$.
Finally,
it should be remarked that (\ref{eq:Ts3}) and (\ref{eq:Ts4})
are stationary with respect to changes of the potential as a result of
standard first order perturbation theory.

The essential problem for obtaining precise total energies at low computational
cost is the incompatibility of the second term
in (\ref{eq:GEtot}) with the second term in (\ref{eq:Ts3}).
These terms differ by
\begin{equation}
\label{eq:DEtot0}
\Delta \Etot =
\sum_n^{} \int_n r^2 \rmd r \sum_{LL'}^{} V^n_{LL'} (r)
n^n_{L'L} (r)
- \sum_n^{} \int_n r^2 \rmd r \sum_{LL'}^{\lmax} \Vproj^n_{LL'} (r)
\nproj^n_{L'L} (r)
\end{equation}
which is the main error made in the evaluation the total energy
if a finite number of potential matrix elements is used. If $\Delta \Etot$
is neglected, the stationarity condition (\ref{eq:Veff}) is violated and
a large number of potential matrix elements must be used if 
precise total energies are desired. If, however, $\Delta \Etot$ is properly
taken into account, much less computational resources are needed as the
numerical investigation in the next section illustrates.

\section{Numerical investigation}
\label{sec:num}

In \cite{ref:Z13} it was found
for the face-centred-cubic (fcc) metals Al, Cu and Pd 
that excellent total-energy results are obtained if
the calculated total energies are corrected according to
the single-particle expression
\begin{equation}
\label{eq:DEtot}
\Delta \Etot = \int_{-\infty}^{\EF} \deps \epsilon n(\epsilon)
- \int_{-\infty}^{\EF} \deps \epsilon \nproj(\epsilon) ,
\end{equation}
which is consistent with (\ref{eq:DEtot0}) because
both (\ref{eq:Ts3}) and (\ref{eq:Ts4}) are stationary with respect to the
potentials. In these calculations with
the screened Korringa-Kohn-Rostoker method \cite{ref:ZDU95, ref:PZD02}
the effective potential was expanded in spherical
harmonics as
\begin{equation}
\label{eq:Vsc}
V(\rv) = \sum_{L}^{\lpot} V_L(r) Y_L(\rvh)
\end{equation}
using $\lpot = 16$. In principle, according to (\ref{eq:VLL0}) this leads to
an infinite number of matrix elements which are obtained by
\begin{equation}
\label{eq:VLLsc}
V_{LL'} (r) = \sum_{L''}^{\lpot} C_{LL'L''} V_L(r)
\end{equation}
where $C_{LL'L''} = \int \drvh Y_L(\rvh) Y_{L'} (\rvh) Y_{L''} (\rvh)$ are
Gaunt coefficients. In the self-consistency steps
a finite number of the matrix elements (\ref{eq:VLLsc}) was used which
is equivalent to the use of non-local projection potentials.
The total energies calculated in this way for different choices of $\lmax$
showed large deviations
from the reference energies obtained with $\lmax = 8$. The deviations
were as large as dozens of millielectron-volts for the standard values
$\lmax=3$ and $\lmax=4$. If, however, a correction as given in
(\ref{eq:DEtot}) was applied, the
deviations of the corrected total energies from the reference energies
amount to about one millielectron-volt or less.

In the present investigation a disordered CrFeCoNi alloy is considered because it is
a more complex system than the fcc systems studied in \cite{ref:Z13}. Whereas
many matrix elements $V_{LL'}$ vanish for the highly symmetric fcc systems,
all matrix elements are non-zero for CrFeCoNi. This system
is a so-called high-entropy alloy, which belongs to a relatively new
class of materials which are technologically important because they can show 
high hardness, wear, oxidation and corrosion resistance.
In a recent large-scale study with up 1372 atoms per unit cell it was found 
\cite{ref:FKS17} that CrFeCoNi tends to form an L${}_{12}$ structure with 
Cr atoms ordered on one fcc sublattice and 
Fe, Co, and Ni atoms randomly distributed on the three other sublattices. For
the present investigation this structure was simulated 
by periodically repeating simple-cubic unit cells with 32 atoms per cell.

\begin{figure}
\begin{center}
\includegraphics[width=.7\textwidth,clip]{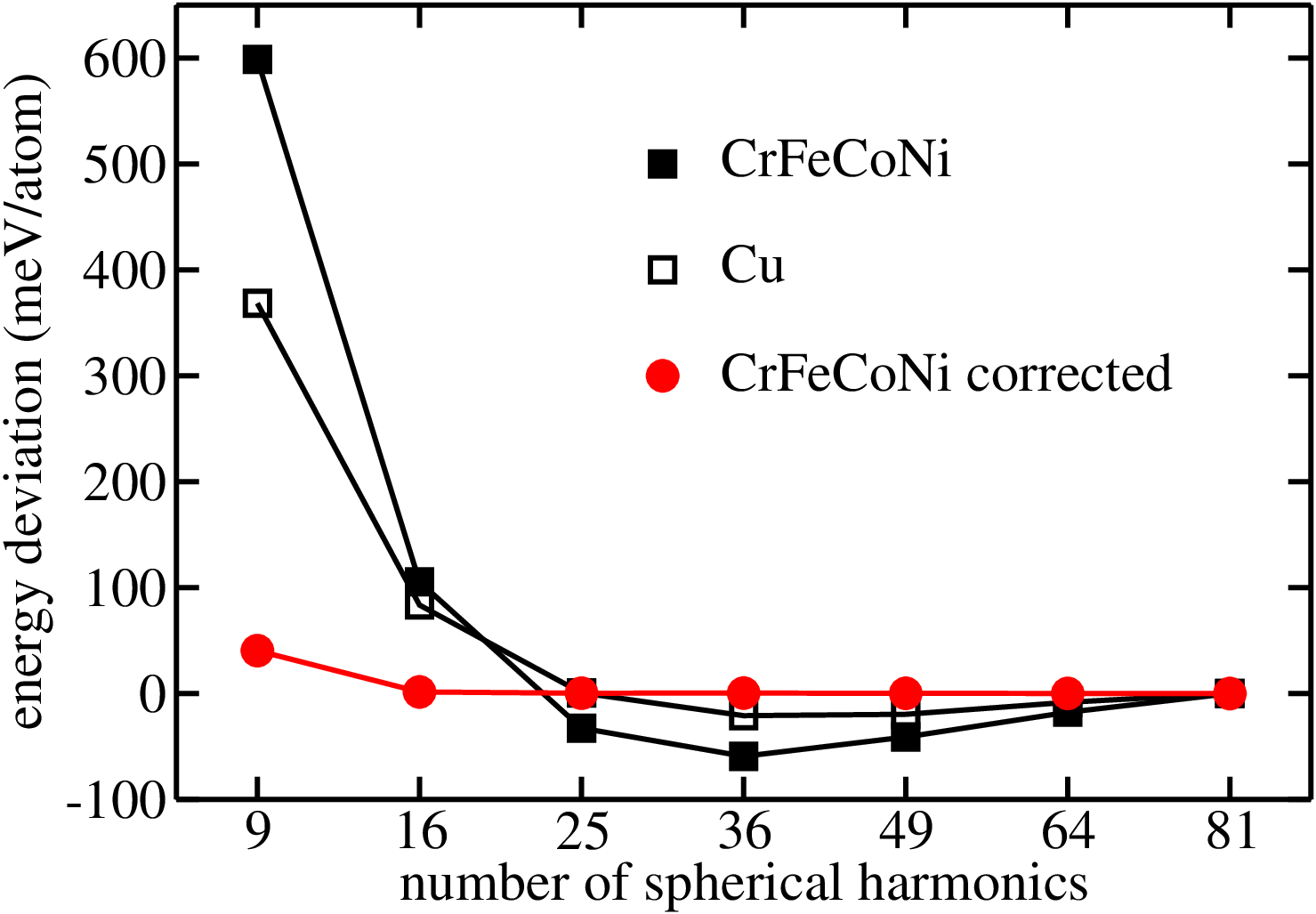}
\caption{\label{fig1}
Total-energy deviations from the reference value for different
numbers of spherical harmonics used for the projection potential.
Full and open squares are for CrFeCoNi and Cu. The results shown as circles have
been calculated for CrFeCoNi by a post-processing procedure
after the self-consistency steps as described
in the text.
}
\end{center}
\end{figure}

As in \cite{ref:Z13}, different choices of
$\lmax$ were used for the projection potentials defined in 
(\ref{eq:V1}), while all other computational parameters
were kept fixed. At the end of the converged self-consistency steps 
the calculated effective potential, the density, the density of states and the total
energy depend on the parameter $\lmax$.
The total-energy results calculated in this way are shown in figure \ref{fig1} as
full squares for different numbers $\Nsph = (\lmax+1)^2$ of spherical
harmonics used in (\ref{eq:V1}).
They are given as deviations from the reference result
obtained with $\lmax=8$. The deviations decrease only
slowly with increasing $\Nsph$ and are as large as
about $100$ millielectron-volts per atom and $-30$ millielectron-volts per atom for the
standard values $\lmax=3$ and $\lmax=4$ which correspond to 
projection into subspaces containing $\Nsph = 16$ and $\Nsph = 25$ 
spherical harmonics. Results 
for fcc Cu, shown in figure \ref{fig1} as open squares,
exhibit the same trend and are of similar size, which
indicates that symmetry is of minor importance
for the precision of total energies.

The slow improvement of the results shown by squares in figure \ref{fig1} arises from
the error made by implicitly neglecting that the exact expression (\ref{eq:Ts3}) 
for the kinetic energy contains an interaction term between potential and density
which does not agree with the term $- \int \drv V(\rv) n(\rv)$ in the total
energy functional (\ref{eq:GEtot}). The slow improvement is not caused
by the approximate calculation of the effective 
potential using a restricted number of potential matrix elements 
in the self-consistency steps. This can be seen if the total energy is evaluated
for the different effective potentials in a post-processing step according to
(\ref{eq:Ts4}) where both the density of states and the density are calculated using
all potential matrix elements up to $\lmax=8$. The total energies calculated in this
way are shown in figure \ref{fig1} as circles. They are much better 
than the results shown as squares and deviate
from the reference value by about 1.5 millielectron-volts per atom for
$\Nsph = 16$, corresponding to $\lmax=3$, and
and by less than 0.5 millielectron-volts per atom for larger values of $\Nsph$.
This good agreement achieved in the post-processing step is a direct consequence
of the stationarity of (\ref{eq:Ts4}) with respect to the potential.
First order changes arising from  
different effective potentials, which are determined by using different numbers of potential 
matrix elements in the self-consistency steps, are cancelled because
they equally contribute to both terms in (\ref{eq:Ts4}) and only second order effects
remain, which are of the order of millielectron-volts per atom.

A technical detail should be mentioned here. The results shown by the
circles in figure \ref{fig1} are obtained in a non-self-consistent way by using a
larger number of potential matrix elements $V_{LL'}$ in the post-processing step than in
the self-consistency steps. This leads to densities of states and densities
which, in metallic systems as studied here, do not satisfy the condition of charge neutrality.
To first order the error in the density of states is compensated by
applying a modified single-particle expression
\begin{equation}
\label{eq:Esp}
E_{\mathrm{sp}} = 
\int_{-\infty}^{\EF} \deps \epsilon n(\epsilon) - 
\EF \left[ \int_{\mathrm{cell}} \drv \nr - N \right] ,
\end{equation}
where $N$ is the total number of electrons in the unit cell of the periodic crystal
and the second term
ensures the stationarity of the total-energy
functional also for non-particle-conserving densities \cite{ref:DWZ89},
and the error in density is compensated by renormalizing the density
with the help of ${\mathrm{Im}} G(\rv,\rv;\EF)$.

\subsection{Perturbation theory for the potential}
\label{sec:vpt}

Considerable computing time is saved for the calculation of the results shown as circles
in figure \ref{fig1} because the $V_{LL'}$ components in
(\ref{eq:V1}) are used only for small subspaces of spherical harmonics
during the self-consistency steps and  
a larger subspace is needed only once in the post-processing step.
A drawback of this procedure is that it does not save computer memory.
The memory increases proportionally to
$\Nat^2 \Nsph^2$ where $\Nat$ is the number of atoms in the system.
Because a large value of $\Nsph$ is required in the post-processing 
step, this quadratic increase can represent a bottleneck for systems with many atoms.

\begin{figure}
\begin{center}
\includegraphics[width=.7\textwidth,clip]{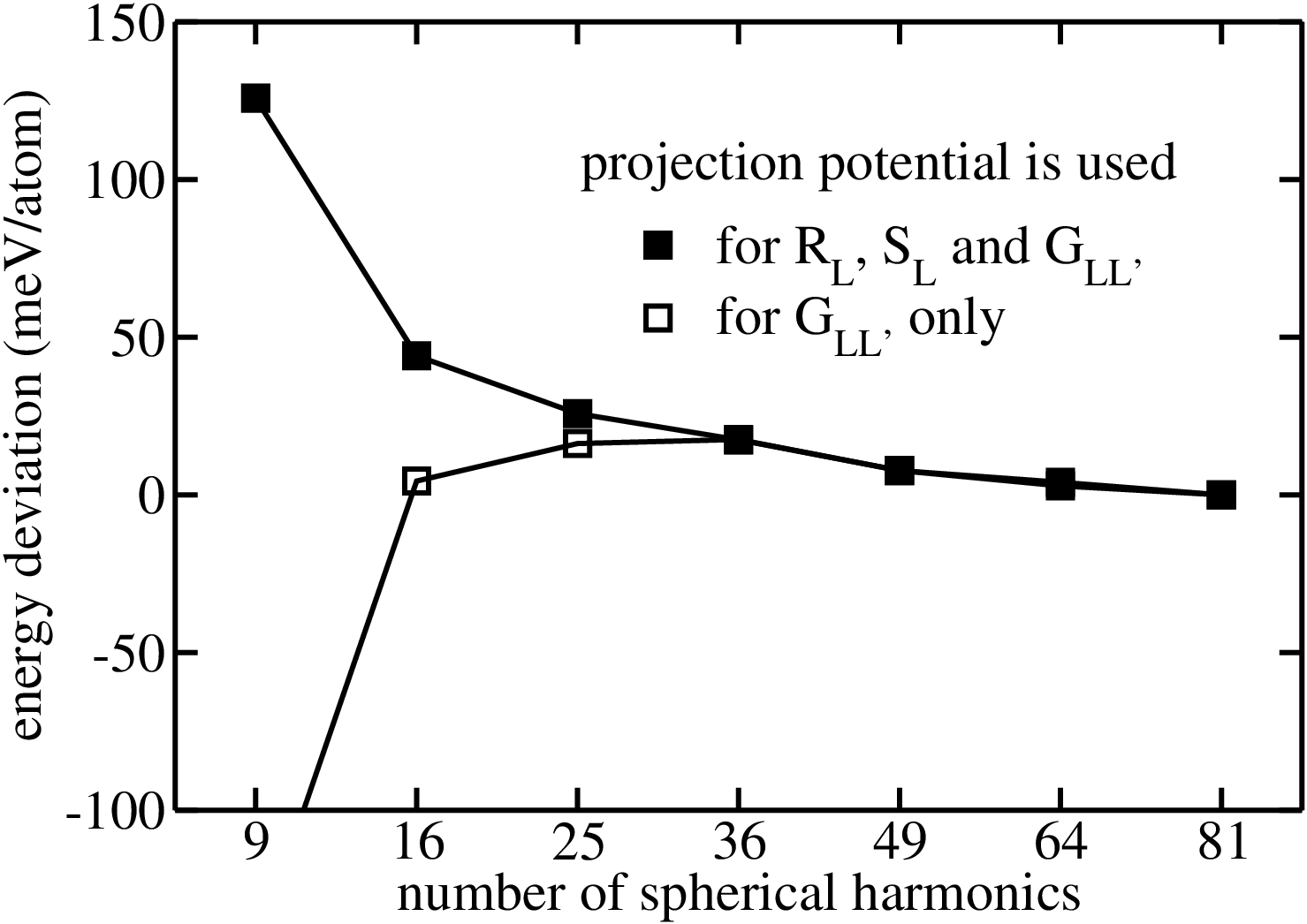}
\caption{\label{fig2}
Total-energy deviations from the reference value for different
numbers of spherical harmonics used for the projection potential.
The results have been obtained by using first order
perturbation theory for the potential. Full and open squares differ by
using the projection potential for the calculation of the functions
$R_L$ and $S_L$ and the Green function matrix elements $G_{LL'}$ or
by using the projection potential only for the calculation of $G_{LL'}$.
}
\end{center}
\end{figure}

Here the question arises whether this bottleneck can be avoided if
$\Delta \Etot$ given by (\ref{eq:DEtot}) is not treated exactly, but in approximations which
take into account only first order effects arising from the different
number of matrix elements in the self-consistency and post-processing steps.
A simple and inexpensive procedure is provided by 
first order perturbation theory which gives the approximation
\begin{equation}
\label{eq:DEtot1}
\Delta \Etot^{(1)} =
\sum_n^{} \int_n r^2 \rmd r \sum_{LL'}^{} \Vproj^n_{LL'} (r)
\nproj^n_{L'L} (r)
- \sum_n^{} \int_n r^2 \rmd r \sum_{LL'}^{\lmax} \Vproj^n_{LL'} (r)
\nproj^n_{L'L} (r) .
\end{equation}
Here in contrast to (\ref{eq:DEtot0}) only the density matrix elements $\nproj_{LL'} (r)$
are needed which are obtained for the effective potentials calculated by using a small value of
$\lmax$ during the self-consistency steps.
Total-energy results corrected with
$\Delta \Etot^{(1)}$ are shown in figure \ref{fig2} by full squares.
Note that vertical axis in figure \ref{fig2} is magnified
by a factor three compared to figure \ref{fig1}. Thus first order
perturbation theory removes about two thirds of the evaluation error which
is nice but not really satisfactory because deviations of about 20
millielectron-volts per atom remain even for $\lmax=6$ corresponding to
$\Nsph =36$. The reason for these still large deviations is that (\ref{eq:DEtot1}) removes
only first order errors arising from the change of the potential but not first order errors 
arising from the change
of the density.

One way to improve the density is to apply the correction (\ref{eq:DEtot1})
in a self-consistent manner. Instead of using (\ref{eq:DEtot1}) in a
post-processing step to update the total energy, it can be used to modify the projection potential
during the self-consistency steps so that
the difference $\Delta \Etot^{(1)}$ vanishes.
This can be done in several ways. For instance, the matrix elements
$\Vproj_{LL'} (r)$ can be adjusted for lower values of
$l$ and $l'$ so that the integrands in both terms
of (\ref{eq:DEtot1}) have equal values for all $r$ or they can be adjusted
in an $r$-independent way by the requirement that both integrals have equal values.
It is also possible to modify only the $l=0$, $l'=0$ component of the projection potential
so that $\Delta \Etot^{(1)}$ vanishes in the self-consistency steps. As the numerical
investigation has shown,
all these modifications lead to minor changes of the calculated total energy which are
less than a millielectron-volt per atom. Thus using (\ref{eq:DEtot1}) to modify the matrix elements 
of the projection potential
does not really improve the calculated total energies, but has the conceptual advantage
that density and total energy are obtained in consistent manner
without the necessity to apply charge-neutrality corrections as discussed above.

Another way to improve the density
at modest increase of computational resources is based on the observation that
the Green function expression (\ref{eq:GF}) contains quantities as 
$R^n_L$, $S^n_L$ and $t_{LL'}$, for which the calculation requires computing times
that increase linearly with $\Nat$, and the Green function matrix elements,
$G^{nn'}_{LL'}$, for which the calculation requires computing times that increase 
cubically with $\Nat$. Thus, for all but the
smallest systems the calculation of $R^n_L$ and $S^n_L$ can be done without $\lmax$ cutoff
and nevertheless the main computational
effort can be saved by restricting
the sums in (\ref{eq:aDyson}) to terms with $l'' \leq \lmax$ and $l''' \leq \lmax$.
The improved calculation of $R^n_L$ and $S^n_L$ 
should lead to better densities, potential and
total energies and to a smaller size of the
first order correction (\ref{eq:DEtot1}).
This is indeed true as the calculated 
open squares in figure \ref{fig2} show, but
only for $\Nsph$ up to $\Nsph = 25$ the improvement is appreciable.
For higher values of $\Nsph$ the deviations are only slightly reduced
which indicates that the second term in (\ref{eq:GF}) dominates the
total-energy evaluation error.

\subsection{Perturbation theory for the density}
\label{sec:npt}

\begin{figure}
\begin{center}
\includegraphics[width=.7\textwidth,clip]{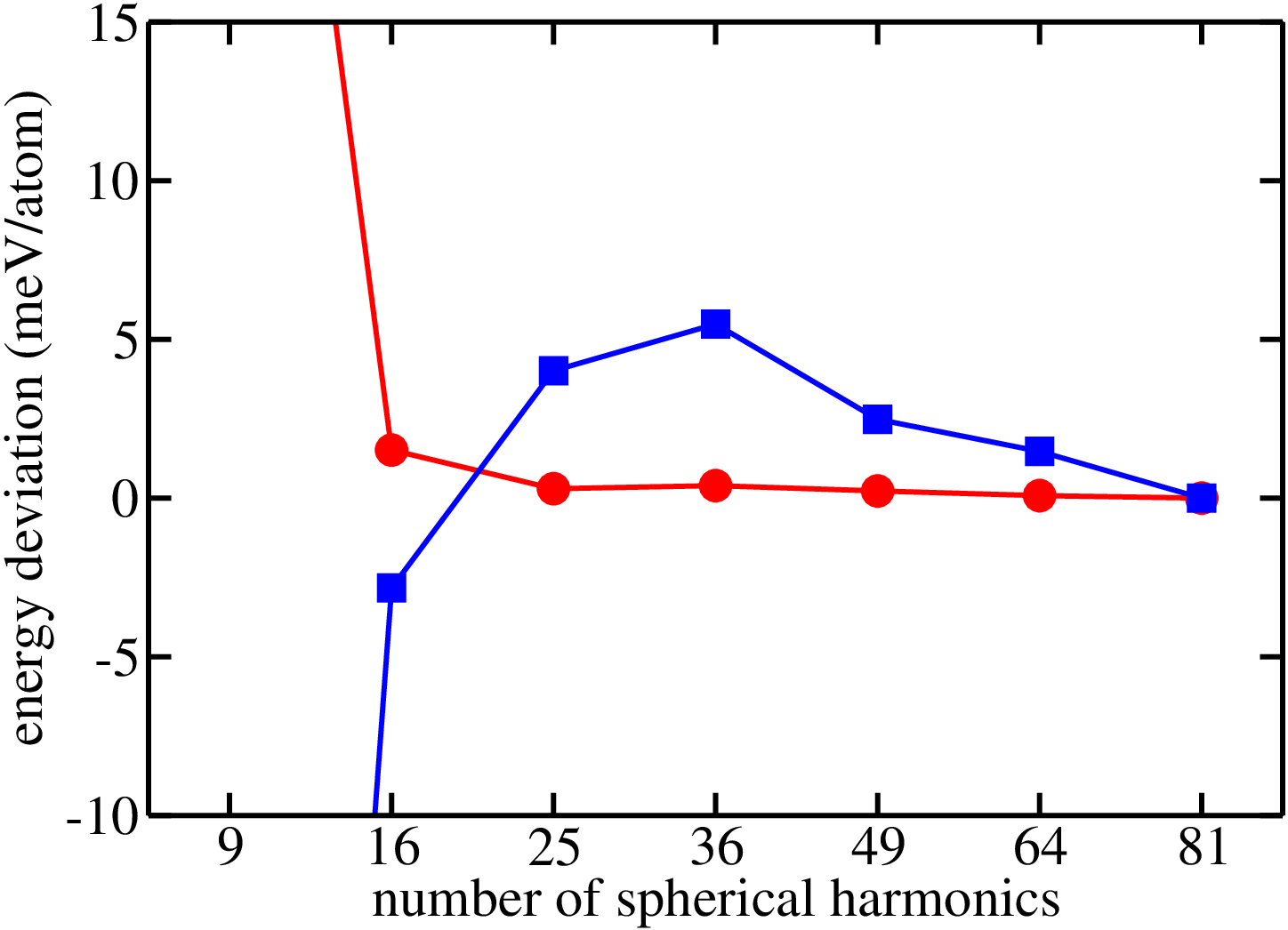}
\caption{\label{fig3}
Total-energy deviations from the reference value for different numbers
of spherical harmonics used for the projection potential.
The results shown as circles have been calculated as in figure \ref{fig1} by a post-processing procedure
and the results shown as squares have been calculated by applying first order perturbation theory
for the density during the self-consistency steps.
}
\end{center}
\end{figure}

While standard first order perturbation theory for the potential
apparently reduces the total-energy evaluation error, it does not provide results
which are comparable in precision to the ones shown as circles in figure \ref{fig1}. The reason
for this is that (\ref{eq:DEtot1}) only compensates the error arising
from the finite sums in (\ref{eq:DEtot0}), 
but not the error which is caused by calculating the density
using potential matrix elements limited by $\lmax$. This limitation can be relaxed
by using more matrix elements in an approximate manner.
In the KKR-GF method this can be done
by recognizing that (\ref{eq:aDyson}) can be solved in two steps. First the
auxiliary Green function matrix elements $\bar{G}^{nn'}_{LL'}$ are calculated
by solving $\bar{G} = g + g t^0 \bar{G}$ and then the matrix elements $G^{nn'}_{LL'}$
are calculated by solving $ G = \bar{G} + \bar{G} \Delta t G$.
Here, $t^0$ has matrix elements limited by $\lmax$ and $\Delta t = t-t^0$
contains the matrix elements neglected in $t^0$. The advantage of
$ G = \bar{G} + \bar{G} \Delta t G$ is that it can be approximated by
$G = \bar{G} + \bar{G} \Delta t \bar{G}$ if the assumption is used that
$\Delta t$ must be treated only to first order. 
Total energies calculated in this way are
shown in figure \ref{fig3} as squares. For $\Nsph \geq 16$ they deviate from the value
for $\Nsph = 81$ only by a few millielectron-volts per atom. Thus, total energies, which are
calculated by using $\Delta t$ in first order perturbation theory, are almost as
good as the ones shown as circles, which are calculated
without this approximation. The big advantage of 
using $ G = \bar{G} + \bar{G} \Delta t \bar{G}$ instead of
$G = \bar{G} + \bar{G} \Delta t G$ is that
only matrix multiplications are needed and that the
computing time is reduced by a factor $\Nat$, because only on-site
($n = n'$) Green function matrix elements are needed to calculate
the density. Thus, total energies can be calculated
essentially by using $\lmax=3$ in the time-consuming parts of the
calculations if deviations of a few millielectron-volts per 
atom are tolerated. Moreover, because the solution of 
$ G = \bar{G} + \bar{G} \Delta t \bar{G}$ is fast, it can be afforded in
each self-consistency step so that effective potential, density and
total energy are obtained in a consistent manner without the need to
apply total-energy or density corrections in a post-processing procedure.

\subsection{Numerical details}
\label{sec:detail}

The calculations were done
with the standard J\"ulich KKR code and partly also with
KKRnano \cite{ref:TZB12} which is a newly developed linear-scaling code suitable for
large systems with up to tens of thousands of atoms.
Both codes are based on the tight-binding (screened) KKR method
\cite{ref:ZDU95, ref:PZD02} and lead to identical results.
The screened structure constants were determined
in real space by using clusters of 55
repulsive muffin-tin potentials with
a constant height of 8 Ryd. The disordered 
CrFeCoNi alloy was described by a simple-cubic unit cell with lattice
constant $712.91$ pm with 32 atoms on ideal fcc positions.
The assumption of ideal positions is justified because the chemical composition is
mainly responsible for disorder effects and additional disorder effects arising from
the small atomic relaxations are not important.
This assumption is also convenient
because highly accurate shape functions can be calculated for the fcc geometry 
\cite{ref:SAZ90}. The core electrons up to 3s and 3p states were
calculated in atomic fashion as described as in \cite{ref:Z13}.
Monkhorst-Pack grids \cite{ref:MP76} were used for the Brillouin-zone integration
with 64 points for CrFeCoNi and
8000 points for Cu.
The Fermi level was chosen at 5 eV by appropriately adjusting
the average value of the potential in the interstitial region outside
the inscribed muffin-tin spheres as discussed in
\cite{ref:Z13}, where also other details, which are not mentioned here, can be found.

\section{Summary and outlook}
\label{sec:summa}

In the present article the problem of calculating precise density-functional total
energies at low cost has been considered from the view point of
non-local projection potentials. It has been discussed that such potentials
can be described exactly by finite numbers of matrix elements in contrast to
local potentials for which an exact description
requires an infinite number of
matrix elements. A consequence of the finite number of matrix elements is that
the orbitals and the density can be calculated practically exactly, which
means that only the approximation of using non-local potentials defined by (\ref{eq:V1}) is decisive for
the precision of the calculated total energies. It has been explained 
that the errors made by approximating the potential
lead to second order errors for the total energy if the generalized
Haydock-Heine total-energy functional is used, but only if
the potential, which appears at two places in the functional, is treated in a 
consistent manner. Unfortunately, this means that a large number of potential matrix elements
must be used for precise calculations of total energies.

For the example of angular projection potentials,
which describe projection into subspaces of spherical harmonics,
it has been demonstrated that precise total energies can be obtained
if the self-consistent density and effective potential are calculated with
a relatively small number of matrix elements and a large number of matrix elements
is used only afterwards to calculate the total energy, 
which leads to considerable savings of computing time.
It also has been demonstrated that, at even lower computational cost,
almost as precise results
can be obtained if the effect of the large number of matrix elements
is taken into account in first order perturbation theory for the density.

The presented analysis of
the Haydock-Heine total-energy functional and the observation
that the usual approximation of the potential by a finite number of matrix elements
is harmful for the stationarity properties of this functional might be
useful in other contexts than the one studied here. For instance, the insight gained
might be used in other ways than by first order perturbation theory.
Furthermore, it might
be used in other electronic-structure methods than the KKR-GF method
by recognizing that the use of a finite number of potential matrix elements can
be understood as working with a non-local potential which arises from projection
into finite subspaces of the chosen basis functions.

\section{Acknowledgments}

The author acknowledges Forschungszentrum J\"ulich for awarding access to
the special purpose computer QPACE3 and acknowledges PRACE for awarding access to
Hazel Hen at GCS@HLRS, Germany.

\section{Appendix}

\subsection{Projection into subspace of plane waves}

In plane wave methods the orbitals are expanded,
in simplified notation without indicating the wave vector $\kv$, as
\begin{equation}
\label{eq:KSpw}
\varphi_i(\rv) = \sum c_i(\Gv) \exp(\rmi \Gv \rv)
\end{equation}
where $\Gv$ are reciprocal lattice vectors and $c_i(\Gv)$
expansion coefficients.
The disadvantage of local potentials is that they
can be described without approximation only by an infinite number of
potential matrix elements
\begin{equation}
\label{eq:Vpw}
V (\Gv,\Gvp) = \int \drv \exp(- \rmi \Gv \rv) V (\rv) \exp(\rmi \Gvp \rv) .
\end{equation}
The neglect of matrix
elements with $|\Gv| > \Gcut$ and $|\Gvp| > \Gcut$ means that
only expansion coefficients $c_i(\Gv)$ with $|\Gv| \leq \Gcut$ are calculated
and only approximate orbitals are obtained.
In contrast to this,
non-local potentials can be defined such that only a finite number of matrix
elements are non-zero. For projection potentials given by
\begin{equation}
\label{eq:Vprojpw}
\Vproj (\rv,\rvp) =
\sum_{\Gv \Gvp}^{|\Gv| \leq \Gcut}
\exp(-\rmi \Gv \rv) \Vproj (\Gv,\Gvp)
\exp(\rmi \Gvp \rvp)
\end{equation}
the matrix elements
\begin{equation}
\label{eq:Vpwnl}
\Vproj (\Gv,\Gvp) = \int \drv \int \drvp
\exp(- \rmi \Gv \rv) \Vproj (\rv,\rvp) \exp(\rmi \Gvp \rv)
\end{equation}
are naturally zero for $|\Gv| > \Gcut$
and for $|\Gvp| > \Gcut$ because of the orthogonality of the plane waves.
Because of that, the orbitals for potentials of type (\ref{eq:Vprojpw})
are exactly given by finite sums and 
the algebraic eigenvalue problem for the determination of the expansion coefficients 
$c_i(\Gv)$ is by construction of finite dimension which means that it can be solved practically
exactly. Thus, except for limiting the sums in (\ref{eq:Vprojpw}) essentially no other
approximation must be made. According to the Haydock-Heine functional the total energy
error should depend only to second order in the potential approximation if errors
arising from evaluation of the functional using two different potentials are avoided.
Similarly as presented in section \ref{sec:num} for projection into
subspaces of spherical harmonics, this might be possible by
using first order perturbation theory for the orbitals and by properly exploiting
the stationarity properties of the total energy functional. Work in this direction
should lead to improved total energies at reduced cost. In fact, an example of such work was
recently published by Canc\`{e}s {\it{et al.}} \cite{ref:CDM16}, where a post-processing
method based on first and second order perturbation theory was presented.

\end{document}